\documentclass[useAMS,usenatbib]{mn2e}
\usepackage{graphicx}
 \usepackage{epstopdf}
\newcommand{\lppr}{\stackrel{<}{\scriptstyle \sim}}

\title[Pulsed VHE emission from the Crab Pulsar] {Pulsed VHE emission from the Crab Pulsar in the context
of magnetocentrifugal particle acceleration.}

\author[Z. Osmanov \& F.M. Rieger]{Z. Osmanov,$^{1}$\thanks{E-mail: z.osmanov@astro-ge.org}
\&  F.M. Rieger$^{2,3}$\thanks{E-mail: frank.rieger@mpi-hd.mpg.de}\\
$^{1}$School of Physics, Free University of Tbilisi, 0183, Tbilisi, Georgia\\
$^{2}$ Zentrum f\"ur Astronomie (ZAH), Institut f\"ur Theoretische Astrophysik, Universit\"at Heidelberg, Philosophenweg 12, 69120 Heidelberg\\
$^{3}$ Max-Planck-Institut f\"ur Kernphysik, P.O. Box 103980, 69029
Heidelberg, Germany }

\begin{document}

\pagerange{\pageref{firstpage}--\pageref{lastpage}} \pubyear{2016}

\maketitle

\label{firstpage}

\begin{abstract}
The Crab Pulsar has been recently detected at very high energies
(VHE) with its pulsed VHE emission reaching up to $1.5$ TeV. The VHE
peaks appear synchronised with the peaks at GeV energies and show
VHE spectra following hard power-law functions. These new findings
have been interpreted as evidence for a gamma-ray production that
happens very close to the light cylinder. Motivated by these
experimental results we consider the efficiency of
magnetocentrifugal particle acceleration in the magnetosphere of the
Crab Pulsar, reexamining and extending results obtained in a
previous work \citep{or09}. It is shown that efficient
magnetocentrifugal acceleration close to the light cylinder could
provide the required electron Lorentz factors of $5\times 10^6$ and
that the resulting inverse Compton (IC) scattering off thermal
photons might explain the enigmatic TeV emission of the pulsar. We
estimate the corresponding VHE luminosity and provide a derivation
of its spectral characteristics, that appear remarkably close to the
observational results to encourage further studies.

\end{abstract}


%%%%%%%%%%%%%%%%%%%%%%%%%%%%%%%%%%%%%%
\section{Introduction}
%%%%%%%%%%%%%%%%%%%%%%%%%%%%%%%%%%%%%%

Rotating neutron stars are today known to be accompanied by pulsed
non-thermal high- and very high energy (VHE) gamma-ray emission
\citep[e.g.,][]{abdo,lem}. While these observations point to the
presence of highly energetic charged particles in the pulsar
environment, the origin of these particles is still an open question
\citep[see e.g.,][]{ahar2,bednarek12,hirotani14,harding15,mochol15}.

According to the standard theory of pulsars, primary particles are
uprooted from the star's surface by means of strong longitudinal
(along the magnetic field line) electrostatic fields \citep{rud} and
efficiently accelerated in a region between the surface and the
light cylinder (LC) area (a hypothetical zone where the linear
velocity of rotation matches the speed of light). In order to
produce VHE emission up to TeV energies, as e.g. seen in the case of
the Crab Pulsar, electrons would need to achieve high Lorentz
factors of the order of $\sim 10^6-10^7$ \citep[e.g.,]
[]{aliu,magic}.

It is still not known which mechanism of acceleration is responsible
for that and where in the magnetosphere this process may take place.
In the framework of e.g. polar cap models longitudinal electrostatic
fields are presumed to allow significant particle acceleration along
open magnetic field lines close to the neutron star
\citep[e.g.,][]{arons,michel}. However, the acceleration efficiency
is usually found to be too strongly limited by curvature and/or
Inverse Compton radiation processes \citep{daugherty,dermer} in
order to account for the production of TeV particles. To overcome
these difficulties, alternative (so-called outer gap) models might
be invoked \citep[e.g.,][]{cheng,chiang,hirotani}, where particles
are accelerated in the outer parts of the magnetosphere. Yet the
anticipated radiation reaction forces along with the relatively
short gap sizes still impose severe constraints on maximum
attainable particle energies. A number of different approaches have
been considered to improve on these problems. \cite{usov} for
example, have examined the intermediate formation of an
electron-positron bound state that could possibly prevent an early
screening of the gap electric field. \cite{muslim}, on the other
hand, have extended the pulsar's electrodynamics by taking into
account general relativistic effects. Although these mechanisms
allowed to slightly increase the gap size and thus led to more
energetic particles, the corresponding energies are still not high
enough when put in the context of recent VHE observations.

In general, the pulsar magnetosphere is characterized by extremely
strong magnetic fields, which in the case of the Crab Pulsar are
believed to be of the order of $4\times 10^{12}$G close to star's
surface. This in turn means that charged particles very efficiently
emit synchrotron radiation with an associated cooling timescale
$t_{syn}\sim\gamma m_0c^2/P_{syn}$ (where $\gamma$, $m_0$ is the
electron's Lorentz factor and mass, respectively, $c$ is the speed
of light, $P_{syn}\approx 2e^4\gamma^2B^2/3m_0^2c^3$ is the single
particle synchrotron power and $e$ is the electron's charge) that is
by many orders of magnitude smaller than the kinematic timescale
$P$, where $P$ is the rotation period, i.e. $P\approx 0.0334$ sec in
the case of the Crab Pulsar \citep[e.g.,][]{lyne}. Hence, charged
particles injected into the magnetosphere will quickly transit to
the ground Landau level and start sliding along the magnetic field
lines. As the field lines co-rotate with the star, the particle
dynamics is then expected to be strongly influenced by the effects
of centrifugal acceleration and the synchrotron losses do not impose
constraints on the maximum attainable energies of particles.

It is clear that these effects might be extremely efficient close to
the LC zone. In the context of the Crab, the possible role of the
centrifugal mechanism to generate the pulsar's emission has been
examined by \cite{gold}. In particular, he has argued that the
co-rotation of plasma particles in the pulsar's magnetosphere could
"create" an efficient energy transfer channel from the rotator into
the kinetic energy of plasma particles. Motivated by this, a
simplified model of centrifugal acceleration was developed by
\cite{macrog}, who considered a rectilinear rotating channel with a
body, freely sliding inside it. In the framework of special
relativity they showed for the first time that the incorporation of
relativistic effects of rotation can lead to radial deceleration,
which is at first sight a rather uncommon phenomenon, yet already
known from general relativity \citep{abram}.

Since then, centrifugal particle acceleration has been studied in a
variety of contexts and used to e.g., predict the location to
frequency mapping in the case of multi-second pulsars
\citep{gang1,gang2} or to analyse the origin of the variable high
energy emission from the rotating jet base in active galactic nuclei
(AGN) \citep[e.g.,][]{ganglesch,riegman,Xu02,orb07,ghis09}. The
results illustrated amongst other the particular role of inverse
Compton scattering on the efficiency of centrifugal acceleration. In
a direct application of centrifugal acceleration to Crab-like
pulsars we have shown that close to the LC area electrons could
achieve Lorentz factor up to $\gamma\sim 10^7$ \citep[][]{or09}. The
analysis of different radiation processes suggested that inverse
Compton scattering in the Crab Pulsar's magnetosphere could lead to
detectable pulsed VHE emission in the TeV band.

Based on the MHD approximation, a detailed investigation of
magneto-centrifugal acceleration of plasma bulk motion close to the
light cylinder has been recently presented by \cite{bog1}, extending
an earlier approach developed in \cite{bog2}. The results are of
particular interest as they lead to very similar expectations in the
case of the Crab Pulsar. In that approach the dynamics of the
particles has been investigated for different 3dim-configuration of
magnetic field lines and it was shown that close to the LC area
particles can achieve extremely high Lorentz factor following a
scaling as previously employed.

The aforementioned works are especially interesting in the context
of the latest results by the MAGIC Collaboration reporting the
detection of pulsed VHE emission up to 1.5 TeV from the Crab Pulsar,
the highest ever detected \citep{magic}. The VHE pulse profiles
revealed two narrow peaks P1 (located at a phase close to $1$) and
P2 (located at a phase close to $0.4$) that appear synchronised with
those seen by Fermi-LAT in the GeV domain. Their detection
significance (in $\sim 320$h of data) is different, though, with
only P2 being significantly detected ($>5\sigma$) at energies above
$400$ GeV. The phase-folded pulse spectra are compatible with
power-law functions over a range of $\sim70$ GeV up to $1.5$ TeV,
with photon indices $\alpha = 3.5 \pm 0.1$ (for P1) and $\alpha =
3.0 \pm 0.1$ (for P2), and TeV flux levels $E^2dN/dEdAdt$ $\lppr
5\times 10^{-14}$ TeV cm$^{-2}$s$^{-1}$ (P1) and $\left(9\pm3\right)
\times 10^{-14}$ TeV cm$^{-2}$s$^{-1}$ (P2), respectively. It is
worth noting that this is the first clear detection of a pulsed
emission component from the Crab in the TeV regime. Unpulsed
(steady) TeV emission, attributed to its nebula, has been detected
quite some time ago \cite[e.g.,][]{weekes89,ahar1} whereas early
evidence for pulsed gamma-ray emission (above a few tens of GeV) has
been related to an inverse Compton (IC) origin in the pulsar wind
zone some tens of light cylinder radii away \citep{ahar2}.

In this paper we reconsider the role of the magneto-centrifugal
acceleration for the Crab Pulsar in the light of the latest VHE
findings, applying and extending results previously obtained
\citep{or09}. We show that the agreement with observations is
successfully close to encourage further modelling. The paper is
organized in the following way. In section~II, basic results of
centrifugal acceleration are recaptured and discussed. Section~III
then presents an application to the Crab Pulsar, while conclusions
are summarised in section~IV.

%%%%%%%%%%%%%%%%%%%%%%%%%%%%%%%%%%%%%%%%%%%
\section[]{Centrifugal acceleration}
%%%%%%%%%%%%%%%%%%%%%%%%%%%%%%%%%%%%%%%%%%%
%
\subsection{Energy-Position Dependence}
In this section we consider the dynamics of centrifugally
accelerated particles close to the LC surface. As noted above,
synchrotron losses will ensure that a charge particle quickly
transits to the ground Landau level and starts to slide along the
field line. Under such conditions it can be shown that the
Lagrangian of a single relativistic massive particle (with rest mass
$m_0$) moving along a co-rotating field line located in the
equatorial plane can be written as \citep{rijmpd}
\begin{equation}\label{L}
L = -m_0\left(1-\upsilon_r^2-\upsilon_{\phi}^2\right)^{1/2},
\end{equation}
where ($c\equiv1$), $\upsilon_r = \dot{r}$ and $\upsilon_{\phi} =
\Omega r+\dot{r}B_{\phi}/B_{r}$ are respectively the radial and
tangential components of velocity, $\Omega = 2\pi/P$ is the pulsar's
angular velocity of rotation, and $B_{r}$ and $B_{\phi}$ are the
corresponding components of the magnetic field, respectively. Since
the Lagrangian is not explicitly time-dependent, the related
Hamiltonian $H=\dot{r} P - L$, where $P=\partial L/\partial \dot{r}$
is the generalized momentum, is a constant of motion (Noether's
theorem). Using Eq.~(\ref{L}) one finds
\begin{equation}\label{H}
H = \gamma m_0\left(1-\Omega r\upsilon_{\phi}\right) = \rm{const},
\end{equation} Accordingly, in the presence of a dominant radial field,
the radial behaviour of the Lorentz factor for a particle injected
at $r_0$ with initial Lorentz factor $\gamma_0$ becomes
\citep{rijmpd}
\begin{equation}\label{gr}
\gamma(r) = \gamma_0\,\frac{(1-r_0^2/r_L^2)}{(1-r^2/r_L^2)},
\end{equation} where $r_L=c/\Omega$ denotes the light cylinder radius.
A charged particle following the field will thus dramatically
increase its Lorentz factor upon approaching the light cylinder
($r\rightarrow r_L$). Using eq.~(\ref{gr}) and the general
definition of $\gamma$, the characteristic acceleration timescale
can be expressed as
\begin{equation}\label{tacc}
t_{\rm acc}(\gamma) = \frac{\gamma}{\dot{\gamma}} \simeq
\frac{\gamma_0^{1/2} (1-r_0^2/r_L^2)^{1/2}r_L}{2 \gamma^{1/2} c}
\propto 1/\sqrt{\gamma}
\end{equation}

In principle, the Lorentz factor dependence, eq.~(\ref{gr}), can
also be obtained within the fluid dynamical framework of the MHD
approximation. As shown recently by \cite{bog1}, the relativistic
dynamics of particles following a rotating field line with a general
3d-shape in a plasma satisfying the frozen-in condition obeys the
equation
\begin{equation}\label{b1}
\frac{1}{\gamma}\frac{\partial\gamma}{\partial R}=\frac{2R}{1-R^2}+
\frac{\upsilon_r\left(\upsilon^2\cos\psi+R\left(1-R^2\right)\frac{\partial\cos\psi}
{\partial R}\right)}{\left(1-R^2\right)\left({\bf
e_re_{_B}}\right)\left(1-\upsilon_d^2\right)},
\end{equation} where $R = r/r_{L}$ is the dimensionless radial coordinate,
$\upsilon$ is the total particle velocity, ${\bf e_r}$ and ${\bf
e_{_B}}$ are the unit vectors along the radial coordinate and along
the direction of the magnetic field line, respectively, $\psi$ is
the angle between ${\bf e_r}$ and ${\bf e_{\phi}}$, ${\bf e_{\phi}}$
is the unit tangential vector, $\upsilon_d = {\bf E\times B}/B^2$ is
the drift velocity and ${\bf E}$ and ${\bf B}$ are the electric and
magnetic field vectors, respectively. As discussed by \cite{bog1},
on approaching the LC ($R = 1$) the second term of the right-hand
side of Eq. (\ref{b1}) remains finite, while the first term
diverges. Therefore, the aforementioned equation reduces to
\begin{equation}\label{b2}
\frac{1}{\gamma}\frac{\partial\gamma}{\partial R}=\frac{2R}{1-R^2},
\end{equation} which no longer depends on the concrete shape of the field line
(as long as it is positively twisted) and has the solution expressed
in eq.~(\ref{gr}). Identifying possible sites in numerical
simulations and adopting parameters characteristic for the Crab
Pulsar, \cite{bog1} found that Lorentz factor up to $ \sim 5\times
10^7$ (at the Alfv\'enic surface) could be achieved by
magnetocentrifugal acceleration.

\subsection{Trajectories - the Archimedean spiral as attractor}
On crossing the light cylinder, magnetic field lines frozen into the
plasma are generally expected to approach an Archimedean spiral
shape with the plasma flowing in it at constant Lorentz factor, i.e.
the Archimedean spiral becomes an attractor \citep{bog1}. This
compares well with the analysis of \cite{grg} who, in an extension
of the work by \cite{macrog}, examined curved rotating channels in
the equatorial plane and studied the relativistic dynamics of a
particle sliding inside this channel. As seen by a laboratory
observer, the effective angular velocity of a particle can be
written as \citep{grg}
\begin{equation}\label{omef}
\Omega_{ef} = \Omega+\phi'(r)\upsilon_r,
\end{equation}
where the shape of the channel is given in polar coordinates in
terms of $\phi$. On the other hand, since we observe plasma
particles escaping the inner magnetosphere, the particle dynamics is
expected to tend to the force-free regime. As a consequence the
effective angular velocity has to vanish eventually and the
particle's radial velocity should saturate. This in turn means that
the physically interesting "trajectories" are those with $\phi(r) =
ar$, where $a$ is a constant. Together with $\Omega$ this explicitly
defines the asymptotic velocity as
\begin{equation}\label{vas}
\upsilon_{as}=-\frac{\Omega}{a}.
\end{equation}
Obviously, if the initial velocity of the particle coincides with
$\upsilon_{as}$, its dynamics will be force-free from the very
beginning. It seems interesting to study what happens if the initial
velocities are different. For this purpose one can consider the
metric tensor on the Archimedean field lines \citep{grg}
\begin{equation}\label{ds}
ds^2 = g_{00}dt^2+2g_{01}dtdr+g_{11}dr^2,
\end{equation} where
\begin{equation}
\label{gab} g_{\alpha\beta} = \left(\begin{array}{ccc}
-\left(1-\Omega^2r^2\right), \;\;\; & a\Omega r
\\ a\Omega r , \;\;\; & 1+a^2r^2 \\
\end{array}\right),
\end{equation}
with indices $\alpha,\beta = {0,1}$ and using $c\equiv 1$. Then, by
introducing the Lagrangian of a particle \citep[e.g.,][]{shapiro}
\begin{equation}
\label{lag} L_A = -m_0
\left(g_{\alpha\beta}\frac{dx^{\alpha}}{d\tau}\frac{dx^{\beta}}{d\tau}\right)^{1/2},
\end{equation} one can see that $t$ is a cyclic parameter for which the associated
conjugate momentum is conserved. Therefore, one can show that the
proper energy of the particle (per unit rest mass) given by
\begin{equation}
\label{ener} E_A =
-\frac{g_{00}+g_{01}\upsilon_r}{\left(-g_{00}-2g_{01}\upsilon_r-
g_{11}\upsilon_r^2\right)^{1/2}},
\end{equation} is a conserved quantity. Here we have employed the fact that based
on the normalisation of the 4-velocity, the Lorentz factor of the
particle in terms of the radial velocity is given by $\gamma(r)
=\left(-g_{00}-2g_{01}\upsilon_r-g_{11} \upsilon_r^2\right)^{-1/2}$.
Equation~(\ref{ener}) then leads to the following relation for the
radial velocity
\begin{eqnarray}
\label{vr} \upsilon_r&=&{\frac{\sqrt{g_{00}+E_A^2}}{g_{01}^2+E_A^2g_{11}}}\nonumber \\
&\times&{\left[-g_{01}\sqrt{g_{00}+E_A^2} \pm E_A
\sqrt{g_{01}^2-g_{00}g_{11}} \right]}.
\end{eqnarray}

% Figure 1: Archimedean Attractor
\begin{figure}
  \resizebox{\hsize}{!}{\includegraphics[angle=0]{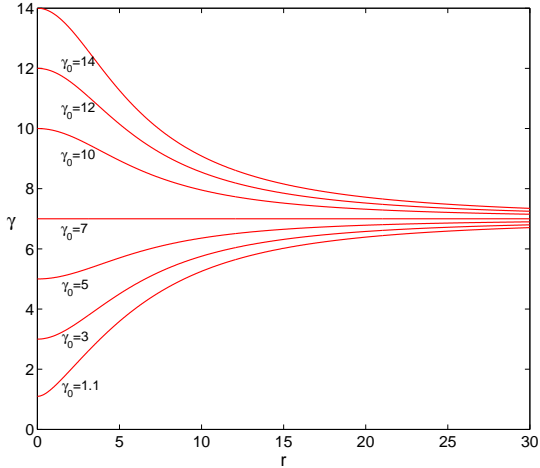}}
  \caption{Characteristic dependence of the Lorentz factor on the radial
  coordinate for different initial Lorentz factors $\gamma_0$. For illustration
  the calculations are done for an Archimedean spiral with small (asymptotic)
  Lorentz factor $\gamma_{as} = 7$. All particles are launched from $r_0 = 0$.
  As it is evident from the plot, the curve $\gamma = \gamma_{as}$ "attracts"
  all other curves.} \label{fig1}
\end{figure}

In Fig. \ref{fig1} the characteristic behaviour of $\gamma(r)$ is
shown for different initial conditions. For illustration and in
order to facilitate a comparison with \cite{bog1}, a small
(asymptotic) Lorentz factor $\gamma_{as} = 7$ has been employed for
the Archimedean spiral. As can be seen, even if the particles are
initially not in the force-free regime, they asymptotically tend to
it. A similar investigation but for 3-dim geometry has been recently
performed by \cite{guda}.

In \cite{bog1} it has been argued that efficient magnetocentrifugal
acceleration occurs close to the Alfv\'enic region (an area where
the Alfv\'enic velocity equals that of plasma particles). By
estimating the corresponding distance in the case of the Crab Pulsar
and using eq.~(\ref{gr}) a maximum attainable value for the Lorentz
factor, $\gamma_{max}\sim 5\times 10^7$ has been inferred. As
discussed in \cite{or09} and considered in more detail in the
following, the aforementioned value could in principle be slightly
reduced by means of energy losses.

%%%%%%%%%%%%%%%%%%%%%%%%%%%%%%%%%%%%%%%%%%%
\section[]{Emission characteristics}
%%%%%%%%%%%%%%%%%%%%%%%%%%%%%%%%%%%%%%%%%%%
In the envisaged scenario, centrifugal particle acceleration can
occur as long as the inertia of the plasma particles does not
counteract efficient co-rotation with the magnetic field. This is
satisfied if the magnetic field energy density $B^2/8\pi$ exceeds
the plasma energy density, $\gamma n_{_{GJ}}Mm_0c^2$, where
$n_{_{GJ}} =\Omega B\cos\alpha/(2\pi ec)$ is the Goldreich-Julian
number density \citep{gj} close to the star, i.e. the number density
of primary particles, $M\approx\gamma/\gamma_0$ is the multiplicity
factor \citep{or09} and $\alpha$ is the inclination angle between
the axis of rotation and the magnetic dipole. For the Crab Pulsar
one has $P=2\pi/\Omega= 0.033$ sec, a nominal light cylinder scale
$r_{L}=c P/2\pi=1.58\times 10^8$ cm, an effective radial distance
$r_l=r_{L}/\sin\alpha$ and a surface magnetic field $B_0\simeq 4
\times 10^{12}$ G. For a dipolar field approximation, the magnetic
field then scales as $B(r_l) \simeq 1.5 \times 10^6\sin^3\alpha$ G
in the LC area. For these parameters, the requirement of efficient
co-rotation leads to a constraint on the maximum Lorentz factor of
\begin{equation}
\label{gcor} \gamma_{max}^{cor}\approx 2 \times 10^7
\left(\frac{\gamma_0}{10^4}\right)^{1/2}\left(\frac{\sin^{3}\alpha}
{\cos\alpha}\right)^{1/2},
\end{equation} where the initial injection Lorentz factor has been normalized on $10^4$
\citep[e.g.,][]{melrose}.

% Figure 2: Magnetosphere-Centrifugal sketch fig2.png
\begin{figure}
 \centering
  \resizebox{5cm}{!}{\includegraphics[angle=0]{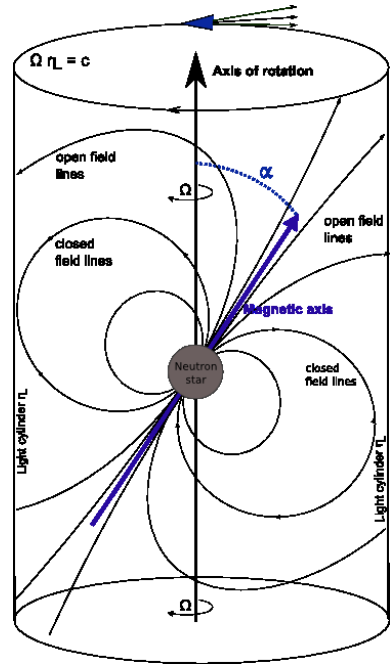}}
  \caption{Sketch of the magnetospheric-centrifugal Pulsar model. Charged particles
  following co-rotating magnetic field lines will be radially decelerated while moving
  outwards but attain a large azimuthal velocity component close to light cylinder (LC)
  surface. A suitably aligned observer will see a strong (beamed) high-energy emission
  pulse for a short fraction of the period.} \label{fig2}
\end{figure}

It is generally assumed that the high energy emission is generated
by particles sliding along open field lines. In the framework of the
present model, these field lines are almost straight in a sense that
their curvature radius exceeds the nominal light cylinder radius.
Yet, in the laboratory frame of reference particles will almost
rigidly rotate on the LC surface \citep{gold}. Therefore, curvature
emission \citep[e.g.,][]{och} could be of significance in limiting
the maximum attainable energies. For Lorentz factor of the order of
$\gamma \sim 10^7$ the characteristic peak frequency,
$\nu_{cur}\approx 3c\gamma^3/(4\pi R_c)$, with $R_c\simeq r_L$, in
the case of the Crab Pulsar becomes
%\citep{or09}
\begin{equation}
\label{nuc} \nu_{cur}\approx 5\times
10^{22}\times\left(\frac{\gamma}{10^7}\right)^3
                              ~{\rm Hz},
\end{equation} which corresponds to photon energies of the order of $0.2
\left(\gamma/10^7\right)^3$ GeV. Hence curvature emission could in
principle lead to a detectable, pulsed GeV contribution. Note that
in principle some mild variation in curvature radii $R_c$ may occur
that could be determined within a more detailed approach
\citep[e.g.,][]{gang2}. We may estimate achievable energies in the
presence of curvature losses by equating the associated cooling time
scale $t_{cur} = \gamma mc^2/P_{cur}$, where $P_{cur}
=2e^2c\gamma^4/(3r_L^2)$ is the single particle curvature energy
loss rate, i.e.
\begin{equation}
\label{tc} t_{cur}\approx 4.5\times
10^{18}\times\frac{1}{\gamma^3}~{\rm sec},
\end{equation} with the acceleration timescale (i.e., eq.~[\ref{tacc}] with $r_L$
replaced by $r_l$) as applied to the Crab Pulsar,
\begin{equation}
\label{tac} t_{\rm acc}\approx \frac{2.6\times
10^{-3}}{\sin\alpha}\times\left(\frac{\gamma_0}{\gamma}\right)^{1/2}
~{\rm sec},
\end{equation} yielding maximum attainable Lorentz factor
\begin{equation}
\label{gc} \gamma_{max}^{cur}\approx 5\times 10^7\sin^{2/5}\alpha,
\end{equation} where we have employed $\gamma_0\approx 10^4$.
Since $\gamma_{max}^{cor}<\gamma_{max}^{cur}$, the relevant
constraint is thus essentially imposed by the requirement of
co-rotation and not by curvature losses. In the approach described
here, efficient particle acceleration is thus taking place in the LC
zone, the energisation being mediated by centrifugal effects and not
magnetic field-aligned electric fields \citep[cf.
also][]{hirotani14}.

In the considered framework, the high energy radiation is generated
in a thin shell of thickness
\begin{equation}\label{d}
d\sim \gamma_0 r_L/\gamma
\end{equation} close to the LC surface \citep{or09}. For a quasi-monoenergetic
particle distribution the total luminosity could be approximated by
$L_{cur}^{GeV} \sim 2 n_{_{GJ}} M\Delta VP_{cur}$, where $\Delta
V\sim \chi \left(\delta l\right)^2 d$ is the corresponding volume,
$\chi\lppr1$ is a dimensionless factor depending on the topology of
magnetic field lines, $\delta l \sim r_L \theta$ is the azimuthal
length scale involved in this process, $\theta \sim \Omega P/10$ is
the corresponding angle, where $P/10$ is an approximate value for
the pulse duration inferred from the light curve observed
\citep{magic}, and where we have taken into account that the pulsar
emits in two channels. Combining the noted values, the pulsed
luminosity generated due to curvature emission at energy
$h\nu_{cur}$ in the GeV band would be of the order of
\begin{equation}
\label{Lc} L_{cur}^{GeV}\simeq 2\times 10^{34}
\left(\frac{\gamma}{10^7}\right)^4
                    \chi\cos\alpha\,\sin^3\alpha\;~{\rm erg/sec}\,.
\end{equation} The specific amount and the spectral shape will depend on the real
(space-energy) distribution of particles and the range of curvature
radii involved, and will need detailed modelling. Given the
magnitude of eq.~(\ref{Lc}) and the fact that the emission is
expected to be significantly focused (beamed; see also below),
curvature radiation is likely to be of prime relevance for
understanding the origin of the $\gamma$-ray emission as seen by
Fermi-LAT \citep{abdo0,magic}.

Given the generic constraints for curvature emission (e.g.,
eq.~[\ref{nuc}]), it is usually believed that inverse Compton (IC)
scattering is responsible for the VHE emission in millisecond
pulsars \cite[e.g.,][]{lyutikov13,harding15,magic}. To this, in
principle a variety of soft photons fields could contribute, e.g.
the thermal emission of the neutron star or secondary synchrotron
emission. As shown in \cite{or09}, in the centrifugal-type approach
following \cite{gold}, the contribution of the star's thermal photon
field to the IC process becomes of relevance as the particle-soft
photon interaction angle is not negligible but of order $\sim
\pi/2$. For the Crab pulsar, the surface temperature is of the order
of $T \simeq 1.2\times 10^6$ K \citep[e.g.,][]{weisskopf},
corresponding to photon energies of $2.8 kT\sim 0.3$ keV. IC
up-scattering of thermal photons to the VHE regime will thus occur
in the Klein-Nishina regime. One could verify that the associated
electron inverse Compton losses will not impose further restrictions
on the maximum attainable energies by recourse to the single
particle emission power \citep[e.g.,][]{blum}
\begin{equation}
\label{PKN}
    P_{KN}\simeq\frac{\sigma_T\left(mckT\right)^2}{16\hbar^3}\left(\ln
                \frac{4\gamma kT}{mc^2}-1.981\right)\left(\frac{r_s}{r_l}\right)^2\,,
\end{equation} where $\sigma_T\approx 6.65\times 10^{-25}$cm$^2$ is the Thomson
cross section, $\hbar\approx 1.05\times 10^{-27}$ erg-sec is the
Planck constant, $k= 1.38\times 10^{-16}$ erg/deg(K) is the
Boltzmann constant and $r_s\approx 11.5$ km is the neutron star's
radius \citep{lat}. To first order the corresponding cooling
timescale $t_{_{IC}}=\gamma mc^2/P_{KN}$ thus increases and becomes
less constraining with Lorentz factor $\gamma$, whereas in contrast
the acceleration timescale, eq.~(\ref{tacc}), decreases as
$1/\gamma^{1/2}$. The noted IC scattering is thus not expected to
impose constraints on the acceleration efficiency. Up-scattering of
thermal photons could, however, lead to pulsed $\gamma$-ray emission
in the VHE domain as pointed out in \cite{or09}. The scattered
photons could well reach energies of the order of $\epsilon\approx
\gamma mc^2\sim 5\left(\gamma/10^7 \right)$ TeV. The MAGIC
experiment has recently reported pulsed VHE emission with an energy
flux level for the higher peak P2 of $F=\left(9\pm 3\right)\times
10^{-14}$TeV cm$^{-2}$s$^{-1}$ in the interval
$\left[965,1497\right]$ GeV \citep{magic}, corresponding to an
isotropic equivalent (spectral) luminosity of $L_{TeV} \simeq
(5-10)\times 10^{31}$ erg/s at a distance of $d\simeq 2$ kpc. In the
considered framework, this emission would be generated by
relativistic particles with Lorentz factors $\sim (2-3)\times 10^6$.
We may again roughly estimate the possible order of magnitude for
the spectral IC luminosity at $\sim1$ TeV expected in the
centrifugal model by multiplying the single particle power $P_{KN}$,
eq.~(\ref{PKN}), with the relevant particle number $N\simeq 2n_{GJ}
M \Delta V$ (two channels) to obtain
\begin{eqnarray}
 L_{IC}^{TeV}\simeq 1.4\times 10^{31} \left(\frac{B}{10^6\rm{G}}\right)
                      \left(\frac{T}{1.2\times 10^6\rm{K}}\right)^2 \chi \cos\alpha \sin^2\alpha
                      ~\frac{\rm{erg}}{\rm{s}}\,.\hspace*{-1cm}\nonumber\\ \label{LTEV}
\end{eqnarray}
One could use this expression to evaluate the degree of focusing and
estimate a fiducial solid angle for the corresponding IC emission
cone of $\Delta \Omega \sim L_{IC}^{TeV}/(d^2F) \sim 2.6
\chi\sin^2\alpha \cos\alpha$ sr if a dipolar field scaling $B(r_l)
\propto \sin^3 \alpha$ close to the LC surface is employed. In
Fig.~\ref{fig2} a sketch of the magnetospheric-centrifugal Pulsar
model is shown. Following the approach developed by \cite{gold}
centrifugally accelerated particles in the LC zone have large
azimuthal velocities and thus approximately emit along a tangent to
the LC surface. It seems evident that in this approach the peaks in
the light curves are in phase for all energy regimes that are
rotationally produced sufficiently close to the LC area.

\begin{figure}
  \resizebox{\hsize}{!}{\includegraphics[angle=0]{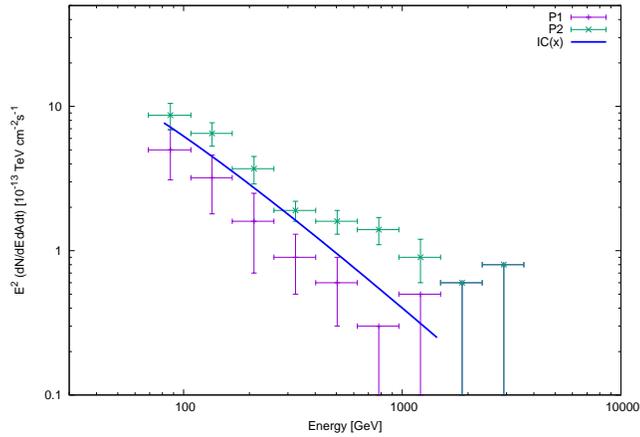}}
  \caption{Illustration of the spectral VHE characteristics expected in the
  magneto-centrifugal Pulsar model as a result of inverse Compton scattering
  of thermal photons in the KN regime. The spectral points are the MAGIC
  (phase-folded) VHE measurements for the main pulse P1 and the interpulse
  P2, respectively \citep{magic}.} \label{fig3}
\end{figure}

\noindent A priori, the radiating particle distribution is not
expected to be mono-energetic, but instead to have some distribution
in energy. This is a complicated issue and, amongst others, likely
to be sensitive on the magnetic field topology, the injection
spectrum and radiation reaction effects. Nevertheless, one can still
evaluate some simple characteristics: Suppose electrons are injected
(by inner gap-type processes) into the magnetosphere at a rate $Q$
with some Lorentz factor $\gamma_0$ and centrifugally accelerated
along a magnetic field line until co-rotation breaks down and they
escape further acceleration by crossing the LC surface. Then to
first order, the differential particle density distribution
$n(\gamma)$ along a field line in the steady state can be
approximated by \citep{riegaha08}
\begin{equation}\label{neq}
n(\gamma) \simeq \frac{Q\,t_{acc}}{\gamma}~ H(\gamma-\gamma_0)
                  \propto \gamma^{-3/2}
\end{equation} using that $t_{acc} \propto \gamma^{-1/2}$, eq.~(\ref{tacc}). Particles
with higher $\gamma$ are located closer to the LC surface and occupy
a thinner shell of width $d \propto \gamma_0/\gamma$, eq.~(\ref{d}),
compared to those at lower energies. An observer will thus see the
integrated emission produced by an effective particle number $N=\int
N(\gamma) d\gamma = \int n(\gamma)d\gamma dV = \int n(\gamma) \Delta
V d\gamma$, i.e., by an effective differential number of electrons
$dN_{e}= N_{e}(\gamma)\,d\gamma$ with a power-law distribution
\begin{equation}\label{neff}
N_{e}(\gamma)=n(\gamma)\Delta V \propto \gamma^{-5/2}\,,
\end{equation} noting that $\Delta V \propto d \propto 1/\gamma$. Up-scattering of a
thermal black body distribution by a power law electron distribution
with power index $p$ in the KN limit will produce a photon spectrum
\citep{blum}
\begin{equation}
\frac{dN_{\gamma}}{d\epsilon_{\gamma} dt} \propto
\epsilon_{\gamma}^{-(p+1)} \left(\ln\frac{\epsilon_{\gamma}
kT}{m^2c^4}+1+C(p)\right)
\end{equation} for photon energies $\epsilon_{\gamma} \ll \gamma_{max} m_e c^2$,
where $C(p=2.5)\simeq -2.4$. The corresponding spectral
characteristics resulting from the power law electron distribution
$p=2.5$ of eq.~(\ref{neff}) turns out to be surprisingly close to
the VHE measurements by MAGIC, reporting photon indices $\alpha =
3.5 \pm 0.1$ (for P1) and $\alpha = 3.0 \pm 0.1$ (for P2). This is
illustrated in Fig.~\ref{fig3}. It is likely that modifications in,
for example, the magnetic field topology and a detailed treatment of
the particle particle injection and transport, aberration and travel
time effects will lead to some index variation around this main
spectral value but full modelling would be required to infer its
plausible range.

% Figure 3: Crab VHE spectrum

\section{Conclusions}
The recent detection of pulsed VHE emission from the Crab Pulsar
\citep[][]{magic} has opened new avenues in pulsar research. The
apparent synchronisation of the pulse profile at GeV and TeV
energies suggests a common property such as a very similar location
of production, the transparency for $\gamma$-ray photons hints to an
outer magnetospheric region, while the expected IC origin points to
the presence of very energetic particles with Lorentz factors of at
least $\gamma\sim 5\times 10^6$. Such characteristics are
challenging to accommodate in traditional emission models. In the
present paper we have argued that magneto-centrifugal particle
acceleration may offer an interesting means to account for the noted
characteristics. In this framework, efficient particle acceleration
takes place in the vicinity of the light cylinder, with the
high-energy $\gamma$-ray emission related to curvature radiation and
the VHE emission (above some tens of GeV) generated by inverse
Compton (IC) upscattering in the Klein-Nishina regime. The model
assumes the occurrence of pair cascades as conventionally treated to
provide the seed particles for injection. A simple analysis of the
anticipated VHE spectral characteristics suggests a power-law index
remarkably close to that actually observed. In the present model the
slight spectral difference reported for the pulses P1 and P2 in the
VHE regime could be related to two different magnetospheric
locations with varying field topology along with possible variations
in the injection spectrum. A full treatment of the corresponding
particle transport and radiation properties, aberration and travel
time effects seems required, however, to elucidate its most
plausible range and to eventually help us understand particle
acceleration processes in pulsar magnetospheres. Given the promising
potential of magneto-centrifugal particle acceleration for
understanding the origin of the observed VHE emission, this seems a
program worth doing. The current MAGIC results rely on an analysis
that combined many periods ($\sim 320$h, spread from 2007 to 2014)
with different sensitivities and energy thresholds, in which the
pulse $P1$ is detected with a moderate significance level of
$\leq2.8\sigma$. Updated VERITAS results (based on $\sim200$h of
data) find a phase-averaged differential spectrum for the Crab
Pulsar compatible with the MAGIC results, but did not yet manage to
establish pulsed emission above 400 GeV \citep{nguyen}. Further
observations with current and future (CTA) instruments
\citep[e.g.,][]{cta}, establishing a homogeneous data set, and
refined analysis will thus be important to better characterise its
spectral evolution and extension, and thereby eventually help us to
understand the origin of the high-energy emission in young pulsars.

%%%%%%%%%%%%%%%%%%%%%%%%%%%%%%%%%%%%%%%%%%%%%%%%%
\section*{Acknowledgments}
%%%%%%%%%%%%%%%%%%%%%%%%%%%%%%%%%%%%%%%%%%%%%%%%%
The research of ZO was partially supported by the Shota Rustaveli
National Science Foundation grant (N31/49). FRM acknowledges
financial support by a DFG Heisenberg Fellowship (RI 1187/4-1).

\end{document}